\documentclass[11pt]{article}
\usepackage{amssymb,amsmath,amsfonts}
\usepackage{graphicx}
\usepackage{graphics}
\usepackage{eepic,epsfig}

\begin{document}

\title{Coherent radiation from a chain of charged particles \\
on a circular orbit around a dielectric ball }
\author{L.Sh. Grigoryan$^1$, A.H. Mkrtchyan$^1$, S.B. Dabagov$^{2,3}$, A.A.
Saharian$^1$\thanks{%
E-mail: saharian@ysu.am}, \and V.R. Kocharyan$^1$, V.Kh. Kotanjyan$^1$, H.P.
Harutyunyan$^1$, H.F. Khachatryan$^1$ \\
\\
\textit{$^1$Institute of Applied Problems of Physics NAS RA, }\\
\textit{25 Hrachya Nersissyan Str., 0014 Yerevan, Armenia }\vspace{0.3cm}\\
\textit{$^2$NR Nuclear University MEPhI, }\\
\textit{31 Kashirskoe Sh., 115409 Moscow, Russian Federation }\vspace{0.3cm}%
\\
\textit{$^3$INFN Laboratori di Frascati, }\\
\textit{54 Via Enrico Fermi, I-00044 Frascati (Roma), Italy }}
\maketitle

\begin{abstract}
We investigate the spectral and angular distribution of the electromagnetic
radiation from a chain of relativistic charged particles uniformly rotating
along equatorial orbit around a dielectric ball. It is shown that, for weak
absorption in the ball material and under relatively mild conditions on the
distribution of the particles, the radiation intensity at specific rotation
frequencies is essentially stronger than the corresponding radiation for a
chain circulating in free space or in a homogeneous transparent medium with
the same dielectric constant as that for the ball. We determine the values
of parameters of the problem for which the charges in the chain emit
coherently and the radiation intensity on a given harmonic increases in
proportion to the square of the number of emitting charges. We also show
that relative shifts in the particles locations up to 10\% do not destroy
the coherence properties of the radiation. It is demonstrated that the
coherence effects may also dominate in the radiation intensity for chains
with non-equidistant distributions of particles. The numerical results
obtained for different dielectric balls have revealed the emitted radiation
to be in the GHz/THz frequency ranges. The high-power radiation from the
chain is confined near the rotation plane within the angular region
determined by the Cherenkov angle for the velocity of the chain image on the
ball surface. In the special case of an equidistant distribution of charged
particles along the orbit the results of the present paper for angle
integrated frequency distribution of the radiation are in agreement with
those previously obtained by our group. We argue that similar coherence
effects will be present in the radiation from a chain of bunches circulating
around the ball.
\end{abstract}

\bigskip


\bigskip

\section{Introduction}

The interaction of charged particles with matter may essentially influence
the characteristics of radiation processes and gives rise to new types of
processes such as Cherenkov radiation (CR), transition radiation, channeling
radiation, etc. \cite{Zrel70}-\cite{Poty11}. In particular, the wide
applications of synchrotron radiation (SR) in fundamental and applied
sciences and in technology motivate the investigations of the influence of
medium on spectral and angular characteristics of the radiation intensity.
SR from a charge uniformly circulating in a homogeneous medium has been
studied in \cite{Tsyt51}-\cite{Gian99}. It was shown that, under the
Cherenkov condition for the velocity of the charge, as a consequence of the
superposition of CR and SR the radiation features may differ significantly
from those for the radiation in free space.

The interfaces separating two media with different electromagnetic
properties may serve as an additional tool to control the spectral and
angular characteristics of SR. As examples of exactly solvable problems of
that kind, in our previous studies we have investigated the radiation from
charges rotating around/inside dielectric ball and cylinder. Those
investigations were based on the recurrence schemes for evaluation of the
electromagnetic field Green tensors in spherically and cylindrically
symmetric piecewise homogeneous media developed in \cite{Arzu95,Grig95}. It
has been shown that the presence of boundaries leads to additional new
features that are absent in the case of homogeneous media. In particular,
the investigations of SR from a charge moving along a concentric equatorial
orbit around or inside a dielectric ball \cite{Arzu95b,Grig06,Arzu08} have
shown that, under the Cherenkov condition for the charge velocity and the
ball material, the so-called "resonance" radiation is generated. The flux of
the resonant radiation in this case exceeds the corresponding value for the
radiation in homogeneous and transparent medium by an order of magnitude.
The radiation intensity for the cases of non-equatorial or shifted
equatorial circular orbits around a dielectric ball has been investigated in
\cite{Grig06b,Grig14}. Similar features were observed for charges on
circular and helical trajectories around or inside a dielectric cylinder
(see \cite{Saha05}-\cite{Kota18} and references therein). SR from a charge
rotating around a cylindrical grating has been studied in \cite{Saha17}. In
\cite{Saha20} the conditions are specified under which strong narrow peaks
appear in the spectral distribution of CR from a charged particle uniformly
moving parallel to the axis of a dielectric waveguide.

The angular distribution of the high-power resonance radiation from a
relativistic electron revolving around a dielectric ball is examined in \cite%
{Grig20}, while the spectral distribution for a chain of equidistant
electrons on a circular orbit has been studied in \cite{Arzu12}. In the
present paper, we investigate the spectral-angular distribution of the
resonance radiation in the GHz and THz frequency ranges generated by a chain
of uniformly moving particles unevenly distributed on a circular trajectory
around a dielectric ball. The paper is organized as follows. In the next
section we describe the problem setup and present the formula for the number
of the radiated quanta. In section \ref{sec:Numer} numerical examples are
given for fused silica and Teflon balls. Qualitative explanation is provided
for coherence effects in the superposition of the radiations from separate
charges. The main results are summarized in section \ref{sec:Conc}.

\section{Statement of the problem and the number of emitted quanta}

\label{sec:Setup}

\begin{figure}[tbph]
\begin{center}
\epsfig{figure=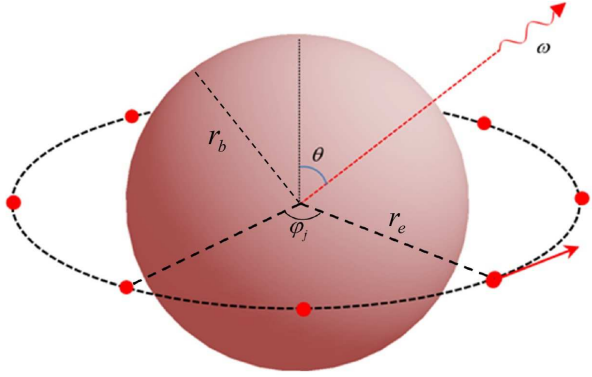,width=8cm,height=5cm}
\end{center}
\caption{A chain of relativistic electrons rotating around a ball in its
equatorial plane. }
\label{fig1}
\end{figure}

We consider a chain of $N$ electrons moving with velocity $v=\mathrm{const}$
along an equatorial circular trajectory around a homogeneous dielectric ball
(see Fig. \ref{fig1}). The radii of the ball and rotation orbit will be
denoted by $r_{b}$\ and $r_{e}$, respectively. In accordance with the
problem symmetry, the spherical coordinate system $(r,\theta ,\varphi )$
will be used with the origin at the ball center and with the polar axis
perpendicular to the rotation plane. Assuming that the chain rotates in the
vacuum, for dielectric permittivity, as a function of the radial coordinate,
one has $\varepsilon (r)=\varepsilon _{b}=\varepsilon _{b}^{\prime
}+i\varepsilon _{b}^{\prime \prime }$ for $r\leq r_{b}$ and $\varepsilon
(r)=1$ in the exterior region $r>r_{b}$. Here, $\varepsilon _{b}^{\prime }$
and $\varepsilon _{b}^{\prime \prime }$ are the real and imaginary parts of
the dielectric permittivity. The magnetic permeability of the ball material
is taken to be equal to unity.

In accordance with the periodicity of the motion the radiation is emitted on
discrete angular frequencies $\omega _{k}=kv/r_{e}$, where $k=1,2,3,\ldots $
determines the number of the radiated harmonic. Let $n_{N}(k,\theta )$ be
the angular density for the number of quanta radiated by the chain on a
given harmonic per rotation period $T=2\pi r_{e}/v$. It can be presented in
the factorized form%
\begin{equation}
n_{N}(k,\theta )=F_{N}(k)n_{1}(k,\theta ),  \label{nN}
\end{equation}%
where $F_{N}(k)$ is the chain structure factor and $n_{1}(k,\theta )$ is the
corresponding quantity for the radiation from a single electron. In the
problem under consideration the factor $F_{N}(k)$ can be written as (see
also \cite{Arzu12})%
\begin{equation}
F_{N}(k)=\left\vert \sum_{j=1}^{N}\exp (-ik\varphi _{j})\right\vert ^{2},
\label{FN}
\end{equation}%
where the angle $\varphi _{j}$ determines the relative position of the $j$%
-th particle inside the chain. The total number of the quanta on the $k$-th
harmonic, emitted per rotation period, is expressed as%
\begin{equation}
n_{N}(k)=\int_{0}^{\pi }n_{N}(k,\theta )d\theta .  \label{nNint}
\end{equation}%
We first analyze the conditions under which the chain generates high-power
radiation at a given harmonic $k$, and then a qualitative explanation of
that phenomenon will be given. We will also determine the dependence of the
effect on random deviations of electrons locations with respect to the
uniform distribution.

For a single electron rotating around a dielectric ball, the formula for the
angular density of the number of quanta radiated per rotation period, $%
n_{1}(k,\theta )=n_{1}(\mathrm{ball},v,x,\varepsilon _{b};k,\theta )$, is
derived in \cite{Grig14}. For the discussion below it is convenient to write
the corresponding expression in a different form given by
\begin{eqnarray}
n_{1}(\mathrm{ball},v,x,\varepsilon _{b};k) &=&32\pi ^{3}\frac{e^{2}}{\hbar c%
}k\sin \theta \left\vert \sum_{s=0}^{\infty }(-1)^{s}\sum_{\mu
=0,1}c_{l}^{(\mu )}\sqrt{\frac{\left( 2l+1\right) ^{2\mu -1}(l-k)!}{%
l(l+1)(l+k)!}}\right.   \notag \\
&&\times \left. \left. \left( \frac{i}{k}\partial _{y}\right) ^{\mu
}P_{l}^{k}(y)\right\vert _{y=0}\mathbf{X}_{l,k}^{(\mu +2)}(\theta
,0)\right\vert ^{2},  \label{n1b}
\end{eqnarray}%
where $l=k+2s+\mu $ and we use the notation $x=r_{b}/r_{e}<1$. In (\ref{n1b}%
), $\mathbf{X}_{l,m}^{(2)}(\theta ,\varphi )$ and $\mathbf{X}%
_{l,m}^{(3)}(\theta ,\varphi )=[\mathbf{n}\times \mathbf{X}%
_{l,m}^{(2)}(\theta ,\varphi )]$ are the spherical vectors of electric and
magnetic types \cite{Bere82}, respectively, and $\mathbf{n}$\ is the unit
vector along the direction determined by the angles $\theta $ and $\varphi $%
. The coefficients $c_{l}^{(\mu )}$ are defined as%
\begin{eqnarray}
c_{l}^{(1)} &=&iu\left[ j_{l}(u)-h_{l}(u)\frac{V_{l}^{j}\left(
xu_{b},xu\right) }{V_{l}^{h}\left( xu_{b},xu\right) }\right] ,  \notag \\
c_{l}^{(0)} &=&(l+1)c_{l-1}^{(1)}-lc_{l+1}^{(1)}+\frac{1-\varepsilon _{b}}{%
x^{2}}\frac{l(l+1)u_{b}j_{l}(xu_{b})}{\left( 2l+1\right) z_{l}(xu_{b},xu)}
\notag \\
&&\times \left[ \sum_{p=\pm 1}\frac{j_{l+p}(xu_{b})}{V_{l+p}^{h}\left(
xu_{b},xu\right) }\right] \left[ \sum_{p=\pm 1}\frac{h_{l+p}(u)}{%
V_{l+p}^{h}\left( xu_{b},xu\right) }\right] ,  \label{cl}
\end{eqnarray}%
with $u=kv/c$ and $u_{b}=kv\sqrt{\varepsilon _{b}}/c$. Here, $j_{l}(y)$ and $%
h_{l}(y)$ are the spherical Bessel and Hankel function (see, for example,
\cite{AbraHand}). For the latter one has $h_{l}(y)=j_{l}(y)+iy_{l}(y)$, with
$y_{l}(y)$ being the spherical Neumann function. In (\ref{cl}) we have used
the notations%
\begin{equation}
V_{l}^{f}\left( xu_{b},xu\right) =j_{l}(xu_{b})\partial
_{x}f_{l}(xu)-f_{l}(xu)\partial _{x}j_{l}(xu_{b}),  \label{Vl}
\end{equation}%
for $f=j$ and $f=h$, and%
\begin{equation}
z_{l}(x,y)=1+\frac{\varepsilon _{b}-1}{2}\sum_{p=\pm 1}\left( 1+\frac{p}{2l+1%
}\right) \left[ 1-\frac{xj_{l}(x)h_{l+p}(y)}{yj_{l+p}(x)h_{l}(y)}\right]
^{-1}.  \label{zlxy}
\end{equation}%
In the absence of the dielectric ball ($\varepsilon _{b}=1$) the formula (%
\ref{n1b}) is reduced to the corresponding expression for SR in free space
(for the radiation in a homogeneous medium see formula (\ref{n1hom}) below).
Note that for given $v$ and $\varepsilon _{b}$, the angular density of the
number of the radiated quanta, given by (\ref{n1b}), is invariant under the
rescaling $r_{b}\rightarrow wr_{b}$, $r_{e}\rightarrow wr_{e}$, $\omega
_{k}\rightarrow \omega _{k}/w$, with dimensionless scaling parameter $w$.

As it has been discussed in \cite{Grig06,Grig20}, under the condition $%
\varepsilon _{b}^{\prime \prime }\ll \varepsilon _{b}^{\prime }$ (weak
absorption of the radiation in the ball material), for specific values of
the parameter $x=r_{b}/r_{e}$ and for large harmonics, $k\gg 1$, the number
of quanta $n_{1}(k)$, emitted by a single charge, significantly exceeds the
number of the radiated quanta in the problems where the same charge is
circulating in vacuum or in a transparent homogeneous medium with dielectric
permittivity $\varepsilon (r)=\varepsilon _{b}^{\prime }$, $0\leq r<\infty $%
. The increase of the radiation intensity is a consequence of coherent
superposition of elementary waves for CR inside the ball emitted near the
charge trajectory and multiply reflected from the ball surface. The fine
tuning of the ratio $x$ is required to ensure the condition for multiple
reflections. The high-power radiation is mainly confined near the rotation
plane in the angular region \cite{Grig20} $\pi /2-\theta _{\mathrm{Ch}}\leq
\theta $ $\leq \pi /2+\theta _{\mathrm{Ch}}$.

The mathematical reason for the appearance of the strong peaks can be
understood as follows. We note that the argument of the spherical Hankel
functions $h_{l}(xu)$, $h_{l+p}(xu)$ in the definition of the function $%
z_{l}(xu_{b},xu)$ is always real and, hence, those functions have imaginary
parts. This means that the function $z_{l}(xu_{b},xu)$ always is imaginary
and has no real zeros with respect to $x$ for real values of the arguments.
This correspond to that the ball has no electromagnetic spherical
eigenmodes. Though the function $z_{l}(xu_{b},xu)$ is always different from
zero, it can be extremely small for large values of $l$. This is based on
the observation that for $0\leq y<1$ and for large $\nu $ the Bessel
function $J_{\nu }(\nu y)$ is exponentially small with respect to the
Neumann function $Y_{\nu }(\nu y)$. This is seen from Debye's asymptotic
expansions \cite{AbraHand} with the leading order term
\begin{equation}
\frac{J_{\nu }(\nu y)}{Y_{\nu }(\nu y)}\sim -\frac{1}{2}e^{-2\nu \zeta
(y)},\;\zeta (y)=\ln \frac{1+\sqrt{1-y^{2}}}{y}-\sqrt{1-y^{2}}.
\label{JYsmall}
\end{equation}%
By using this property, we write $h_{l}(y)=iy_{l}(y)[1-ij_{l}(y)/y_{l}(y)]$
and for large $l$ expand the function (\ref{zlxy}) over the small ratio $%
j_{l}(y)/y_{l}(y)$:
\begin{equation}
z_{l}(x,y)\approx z_{l}^{(0)}(x,y)-i\frac{\varepsilon _{b}-1}{2}\frac{x}{y}%
\sum_{p=\pm 1}\frac{\left( p+\frac{1}{2l+1}\right) j_{l}(x)j_{l+p}(x)}{\left[
yj_{l+p}(x)y_{l}(y)-xj_{l}(x)y_{l+p}(y)\right] ^{2}},  \label{zlexp}
\end{equation}%
with the leading order term%
\begin{equation}
z_{l}^{(0)}(x,y)=1+\frac{\varepsilon _{b}-1}{2}\sum_{p=\pm 1}\left( 1+\frac{p%
}{2l+1}\right) \left[ 1-\frac{xj_{l}(x)y_{l+p}(y)}{yj_{l+p}(x)y_{l}(y)}%
\right] ^{-1}.  \label{zlxy0}
\end{equation}%
Unlike the function $z_{l}(x,y)$, the function $z_{l}^{(0)}(x,y)$ is real
for real values of the arguments and it may become zero. Now from (\ref%
{zlexp}) we see that at those zeros the function $z_{l}(x,y)$ is
exponentially small and gives an exponentially large factor in the
expression (\ref{n1b}) for the angular density of the number of radiated
quanta. This does not yet mean that the angular density will be
exponentially large because the contributions of the other terms should be
estimated as well. An important thing which should be emphasized here is
that the values for the ratio $x$ for which strong peaks are present in the
radiation intensity are determined by the zeros of the function (\ref{zlxy0}%
) with high accuracy. We have checked that by numerical examples. Another
important thing, based on the analytic estimates given above, is that the
peaks come from the electromagnetic field modes of the electric type. The
electromagnetic modes corresponding to the zeros of the function can be
termed as "quasieingemodes" of the ball. They have a large confinement time
inside the ball with multiple reflections from the ball surface (in the
absence of absorption the exact eigenmodes would have an infinite
confinement time).

The arguments presented are similar to those given in \cite{Saha05} for the
radiation from a charge circulating inside or around a dielectric cylinder.
A factor that is the analog of the function (\ref{zlxy}), with the spherical
Bessel functions replaced by the cylindrical functions with an integer
order, is present in the corresponding expression for the radiation
intensity (the factor $\alpha _{m}$, $m=1,2,\ldots $, in \cite{Saha05}). The
Hankel functions appear in the corresponding expression in the form $H_{m\pm
1}(\rho _{1}\sqrt{m^{2}\omega _{0}^{2}\varepsilon _{1}/c^{2}-k_{z}^{2}})$,
where $\rho _{1}$ is the radius of the cylinder, $\omega _{0}$ is the
angular velocity of the charge circulation, $\varepsilon _{1}$ is the
dielectric permittivity of the medium surrounding the cylinder, and $k_{z}$
is the projection of the wave vector on the axis of the cylinder. An
important difference from the case of dielectric ball is that now there are
electromagnetic modes for which the argument of the Hankel functions becomes
purely imaginary (modes with $|k_{z}|>m\omega _{0}\sqrt{\varepsilon _{1}}/c$%
) and they are transformed to Macdonald function. For those modes the
function $\alpha _{m}$ becomes real and it may have zeros with respect to $%
k_{z}$. They are the eigenmodes of the dielectric cylinder. The
corresponding radiation fields are exponentially suppressed in the region
outside the cylinder and they propagate inside the cylinder.

\section{Numerical analysis and qualitative explanation}

\label{sec:Numer}

Below we consider a chain of electrons with energy $E_{e}=2\,\mathrm{MeV}$
rotating around a ball made of quartz or Teflon and generating high-power
radiation in the GHz or THz frequency ranges in dependence of the optimal
choices for the values of the system parameters $N,\varphi
_{j},r_{e},r_{b},k,\varepsilon _{b}$.

\subsection{Chain rotating around a ball of fused quartz}

\label{ssec:Quartz}

We start the discussion of the radiation features from the case of a chain
on a circular orbit around a ball made of fused quartz.

\subsubsection{Radiation from a single electron}

\begin{figure}[tbph]
\begin{center}
\epsfig{figure=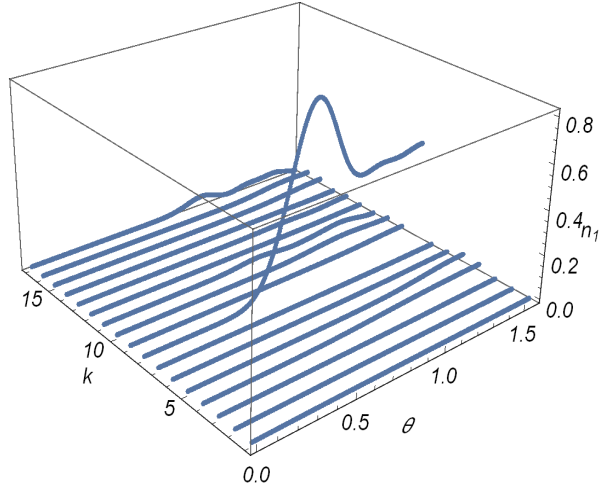,width=8.cm,height=6.5cm}
\end{center}
\caption{The angular distribution of the number of quanta $n_{1}(k,\protect%
\theta )$ generated by a single electron per rotation period as a function
of the harmonic number $k$ and polar angle $\protect\theta $ (in radians).
The graphs are plotted for an electron orbiting in free space around a ball
made of fused quartz.}
\label{fig2}
\end{figure}

\begin{figure}[tbph]
\begin{center}
\begin{tabular}{cc}
\epsfig{figure=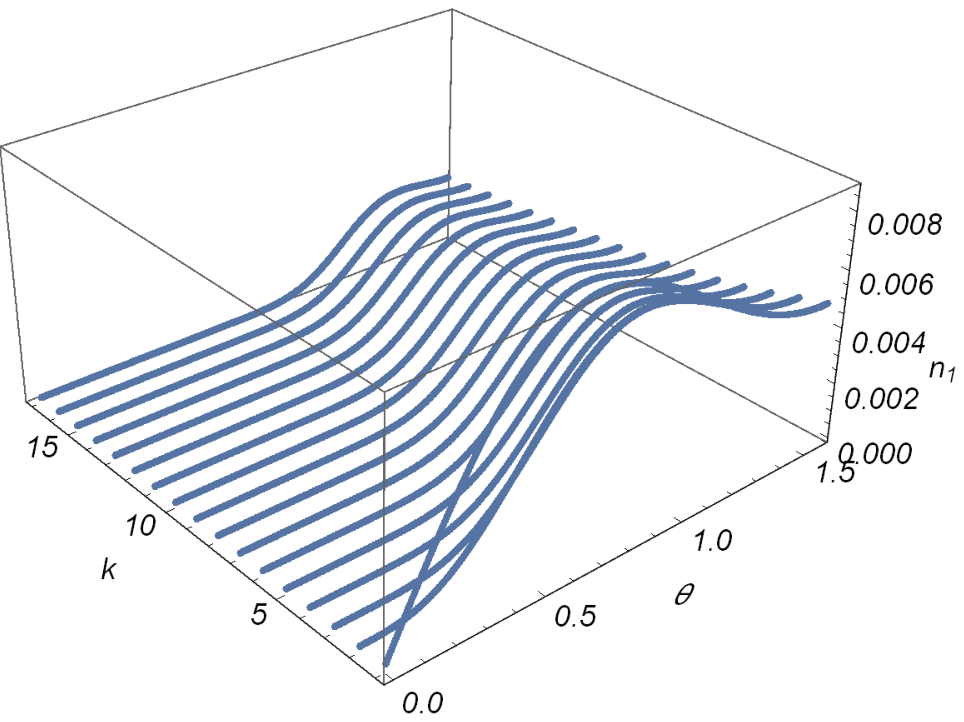,width=7.cm,height=5.5cm} & \quad %
\epsfig{figure=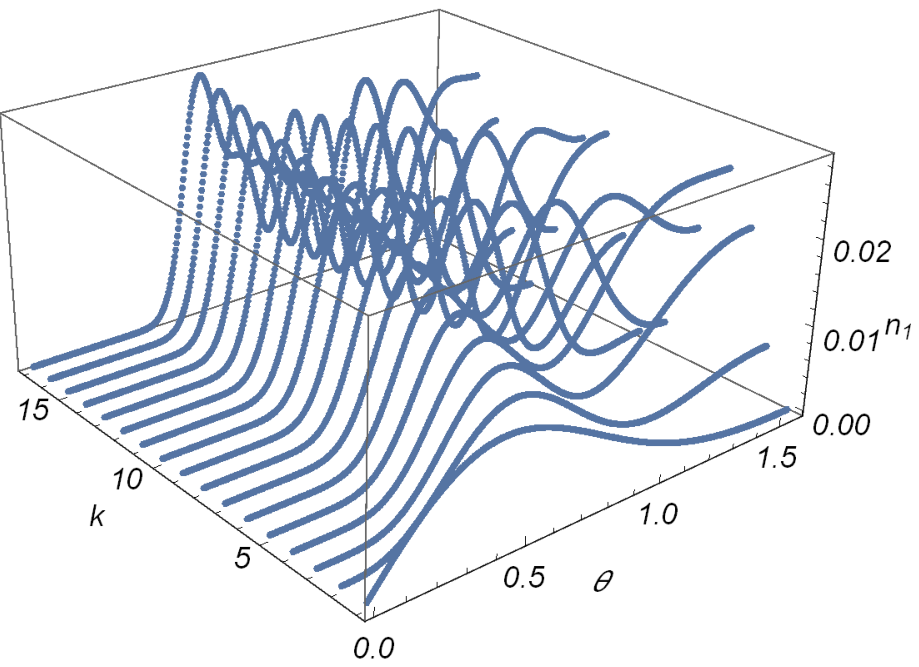,width=7.cm,height=5.5cm}%
\end{tabular}%
\end{center}
\caption{The same as in figure \protect\ref{fig2} for an electron
circulating in vacuum (left panel) and in a homogeneous transparent medium
with dielectric permittivity $\protect\varepsilon =\protect\varepsilon %
_{b}^{\prime }=3.78$ (right panel).}
\label{fig3}
\end{figure}

Let us denote by $v_{\ast }=vx$ the velocity of the particle's image on the
surface of the ball and by $\theta _{\mathrm{Ch}}=\arccos (c/v_{\ast }\sqrt{%
\varepsilon _{b}^{\prime }})$ the related Cherenkov angle, assuming that the
Cherenkov condition $v_{\ast }\sqrt{\varepsilon _{b}^{\prime }}/c>1$ is
satisfied. In Fig. \ref{fig2} we display the results of numerical
calculations for the angular density of the number of quanta, $n_{1}=n_{1}(%
\mathrm{ball},v,x,\varepsilon _{b};k,\theta )$, radiated per rotation period
by a single electron of energy 2 MeV in a circular orbit around a fused
quartz ball with $\varepsilon _{b}=3.78(1+10^{-4}i)$ \cite{Voro65,Chud21}.
For the harmonic $k=8$ the abovementioned amplification effect is obtained
for the value $r_{b}/r_{e}=0.9815$. We have numerically checked that this
value coincides with the zero of the function $z_{l}^{(0)}(xu_{b},xu)$ with
rather good accuracy, in accordance of the analytic arguments given above.
In order to see the influence of the dielectric ball on the radiation
intensity, on the left panel of Fig. \ref{fig3} we have plotted the angular
density of the number of quanta, $n_{1}=n_{1}(\infty ,v,\varepsilon
;k,\theta )$, for an electron rotating in vacuum, $\varepsilon =1$ (the ball
is absent). The right panel of figure \ref{fig3} presents the corresponding
quantity, $n_{1}=n_{1}(\infty ,v,\varepsilon ;k,\theta )$, for the radiation
from a single electron in a homogeneous transparent medium with dielectric
permittivity $\varepsilon =\varepsilon _{b}^{\prime }=3.78$. The graphs in
Fig. \ref{fig3} are plotted by using the formula \cite{Zrel70}%
\begin{equation}
n_{1}(\infty ,v,\varepsilon ;k,\theta )=\frac{2\pi e^{2}k}{\hbar c\sqrt{%
\varepsilon }}\left[ J_{k}^{2}(k\beta \sin \theta )\cot ^{2}\theta +\beta
^{2}J_{k}^{\prime 2}(k\beta \sin \theta )\right] \sin \theta ,  \label{n1hom}
\end{equation}%
for SR in a homogeneous transparent medium with dielectric permittivity $%
\varepsilon $. Here, $\beta =v\sqrt{\varepsilon }/c$ and $J_{k}(x)$ is the
Bessel function. The radiation frequency is given by $\nu _{k}=kv/(2\pi
r_{e})$. For the resonant radiation on the harmonic $k=8$, corresponding to
Fig. \ref{fig2}, and for the ball radius $r_{e}=1\,\mathrm{cm}$ one gets $%
\nu _{8}\approx 37\,\mathrm{GHz}$. The radiation frequency can be tuned by
the choice of the ball radius for a given $r_{e}$ or by the choice of the
latter for a given $r_{b}$.

As illustrated in Figs. \ref{fig2} and \ref{fig3}, for a given value of the
radiation harmonic $k$ (the harmonic $k=8$ in the example at hand) and for
fixed values of the other parameters, by tuning the ratio $r_{b}/r_{e}$ we
can obtain an essential increase of the radiation intensity compared with
the corresponding radiation for an electron circulating in free space or in
a transparent homogeneous medium having dielectric permittivity equal to the
real part of the dielectric permittivity for the ball. We also see an
significant increase of the radiation intensity in a homogeneous transparent
medium compared with the radiation in vacuum.

\subsubsection{Radiation from a chain}

The spectral-angular distribution of the number of quanta radiated by a
chain is given by the formula (\ref{nN}). For equidistant distribution of
electrons in the chain one has $\varphi _{j}=2\pi (j-1)/N$ and the
expression (\ref{FN}) for the factor $F_{N}(k)$ is simplified to (see also
\cite{Arzu12})
\begin{equation}
F_{N}(k)=\left\{
\begin{array}{cc}
N^{2}, & \text{for }k=mN \\
0, & \text{for }k\neq mN%
\end{array}%
\right. ,\;m=1,2,\ldots  \label{FN2}
\end{equation}%
This shows that the radiation takes place on the angular frequencies $\omega
_{m}=Nmv/r_{e}$. Of course, that is natural, because for an equidistant
chain the period of the system motion $T_{N}$ is given by $T_{N}=T/N=2\pi
r_{e}/(Nv)$ and for the radiation frequencies we get $\omega _{m}=2\pi
m/T_{N}$ with $m=1,2,\ldots $. The electrons in the chain radiate coherently
and the radiation intensity is amplified by the factor $N^{2}$.

\begin{figure}[tbph]
\begin{center}
\epsfig{figure=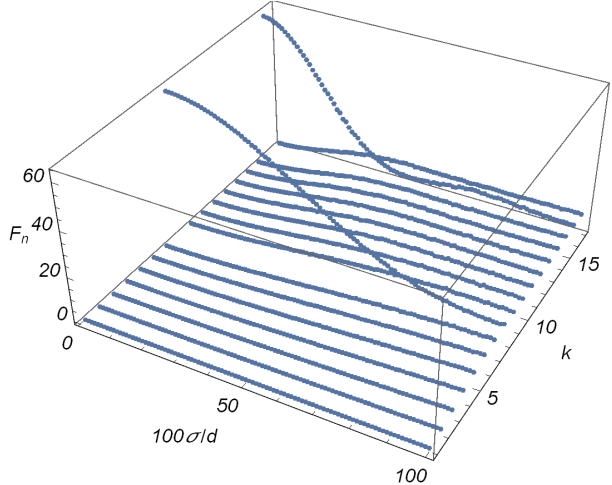,width=8cm,height=6.5cm}
\end{center}
\caption{The structure factor of the chain of $N=8$ electrons versus the
relative maximal shift $\protect\sigma /d$. The shift values are given in
percentages. }
\label{fig4}
\end{figure}

In realistic situations the condition for the same separation between
neighboring particles is obeyed approximately and it is of interest to see
the sensitivity of the coherent superposition considering small variations
in the equidistant distribution. Figure \ref{fig4} shows the dependence of
the chain structure factor on the relative maximal shift $\sigma /d$ for
each particle, where $d=2\pi r_{e}/N$ is the arc distance between
neighboring particles. The shift from the equidistant position is taken as a
randomly distributed quantity with the uniform distribution (of width $%
\sigma $) with the maximal shift equal to the half of the arc distance $d$.
From the data presented in Fig. \ref{fig4} it follows that the chain
radiates coherently, i.e. $F_{N}\sim N^{2}$, for up to 10\% of relative
shifts $\sigma /d$. Thus, we can conclude that the replacement of a single
charge by the chain formed by $N=8$ equidistant charges rotating about the
ball, additionally increases the radiation by a factor $\sim N^{2}$.

\begin{figure}[tbph]
\begin{center}
\epsfig{figure=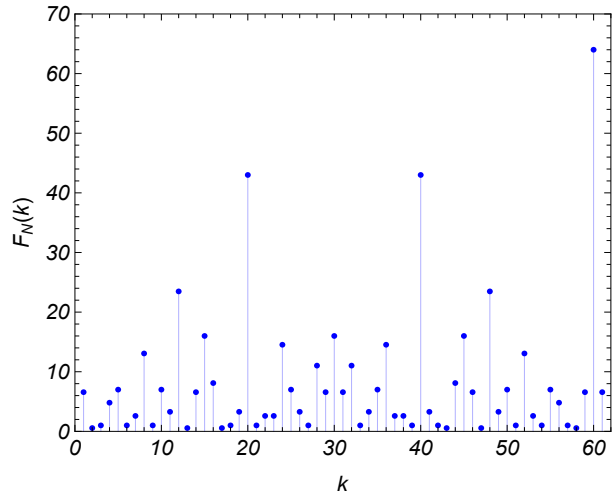,width=8cm,height=6.5cm}
\end{center}
\caption{The structure factor of the chain of $N=8$ particles distributed in
accordance with (\protect\ref{rj}). }
\label{fig5}
\end{figure}

The coherence effect discussed above for the equidistant distribution may
take place for more general distributions of charges on a circular
trajectory. Let us consider the distribution of $N$ charges given by%
\begin{equation}
\varphi _{1}=0,\;\varphi _{j}=2\pi r_{j},\;j=2,\ldots ,N,  \label{dist2}
\end{equation}%
at $t=0$, where $r_{j}$ are positive rational numbers such that $%
r_{2}<r_{3}<\cdots <r_{N}<1$. In the special case of the equidistant
distribution one has $r_{j}=(j-1)/N$. Let $r_{j}=p_{j}/q_{j}$ be the
standard form of the rational number $r_{j}$. For the sum in the structure
factor (\ref{FN}) we get $1+\sum_{j=2}^{N}\exp (-2\pi ikp_{j}/q_{j})$. From
here it follows that for the harmonics of the radiation with $k=mk_{0}$,
where $m$ is a positive integer and $k_{0}$ is the least common multiple of
the numbers $q_{2},q_{3},\ldots ,q_{N}$, the chain radiates coherently with $%
F_{N}(mk_{0})=N^{2}$. However, unlike the case of equidistant distribution,
in general, the radiation on harmonics $k\neq mk_{0}$ is present as well.
The radiation on those harmonics can be of the order $N^{2}$, though $%
F_{N}(k)<N^{2}$. To illustrate this point, in Fig. \ref{fig5} the function $%
F_{N}(k)$ is displayed for the distribution of $N=8$ particles with the
locations described by
\begin{equation}
\{r_{j}\}_{j=2,\ldots ,N}=\left\{ \frac{1}{5},\frac{1}{4},\frac{1}{3},\frac{2%
}{5},\frac{1}{2},\frac{3}{5},\frac{3}{4}\right\} .  \label{rj}
\end{equation}%
For this example $k_{0}=60$. Note that the function $F_{N}(k)$ is periodic
with the period equal to $k_{0}$. We can have a more general case of
distribution when the condition for the coherent superposition is obeyed
only for a subsystem of particles in the chain. In this case one has a
partial coherence with the radiation intensity proportional to the square of
the number of particles in that subsystem.

\subsection{Radiation for a Teflon ball}

\begin{figure}[tbph]
\begin{center}
\begin{tabular}{cc}
\epsfig{figure=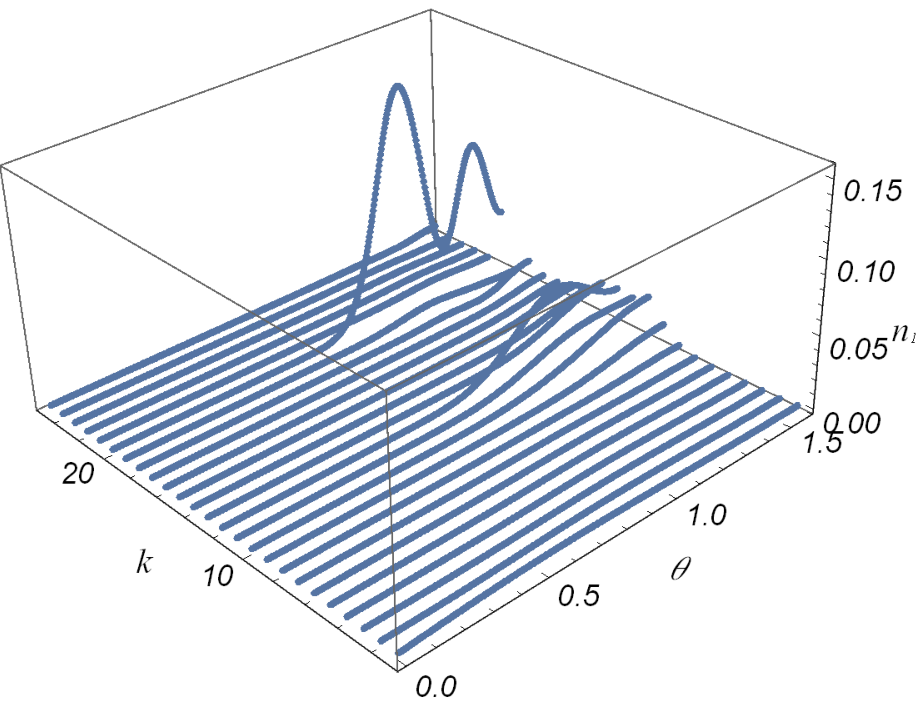,width=7.cm,height=5.5cm} & \quad %
\epsfig{figure=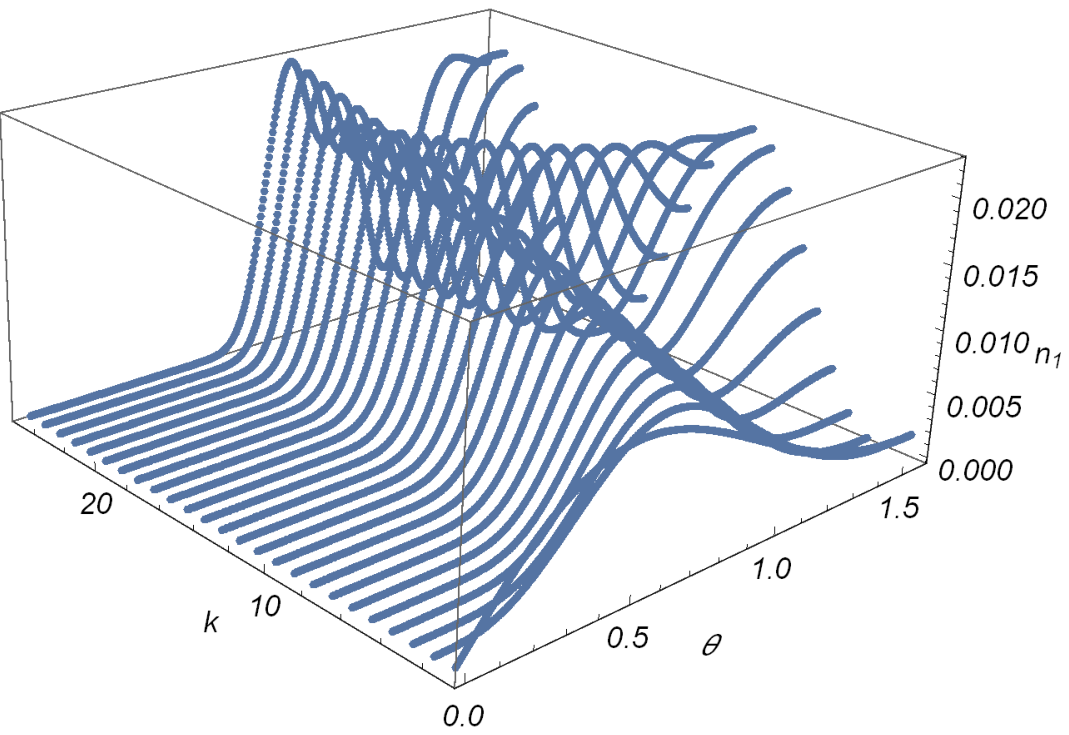,width=7.cm,height=5.5cm}%
\end{tabular}%
\end{center}
\caption{The same as in figure \protect\ref{fig2} for a Teflon ball. The
graphs are plotted for a charge rotating in vacuum around a Teflon ball
(left panel) and in a transparent and homogeneous medium with permittivity $%
\protect\varepsilon =\protect\varepsilon _{b}^{\prime }=2.2$ (right panel).}
\label{fig6}
\end{figure}

\begin{figure}[tbph]
\begin{center}
\epsfig{figure=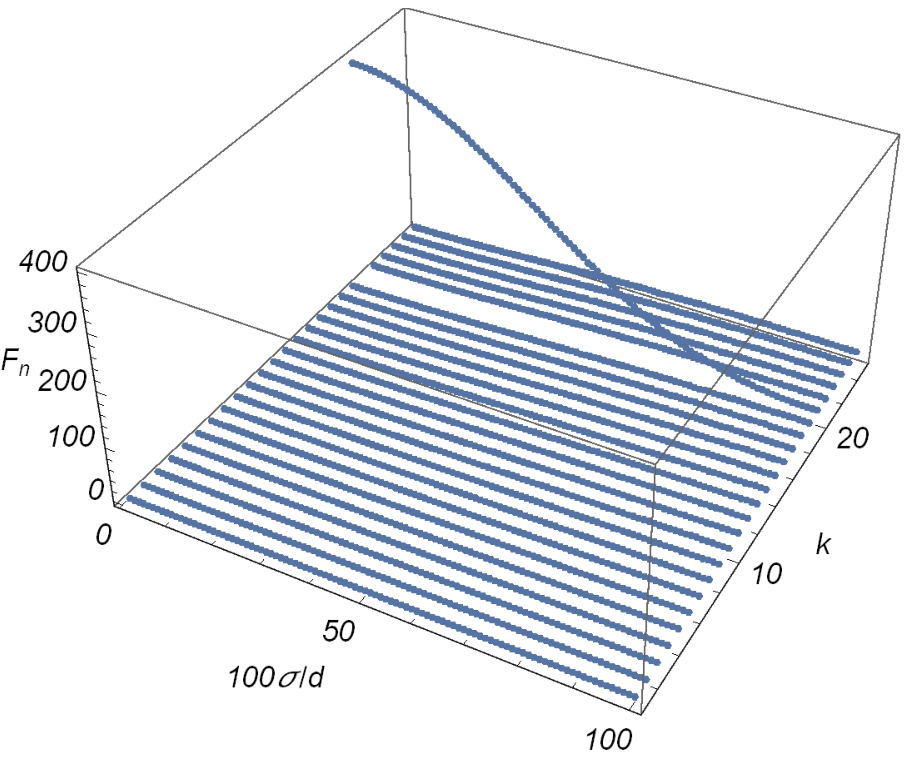,width=8cm,height=6.5cm}
\end{center}
\caption{The same as in figure \protect\ref{fig3} for a chain of $N=20$
particles.}
\label{fig7}
\end{figure}

Similar results are obtained for a ball made of Teflon. In this case the
parameters of the system have the following values: $\varepsilon
_{b}=2.2(1+0.0002i)$, $r_{b}/r_{e}=0.9616$, $k_{0}=20$. The corresponding
graphs are presented in Figs. \ref{fig6} and \ref{fig7}. As one can see from
presented figures, the amplification effect takes place in this case as
well. Taking, for example, $r_{e}=0.15\,\mathrm{cm}$, for the frequency of
the resonance radiation we get $\omega _{k_{0}}/2\pi =6\cdot 10^{11}\mathrm{%
Hz}$ which is in the THz range.

\subsection{Qualitative explanation}

Here we provide a qualitative explanation of the radiation intensity
enhancement. As it already has been mentioned above, compared with the case
of a single particle, for an equidistant distribution of $N$ particles on a
circle the period of motion is reduced by the factor $N$ and, as a
consequence, the radiation frequencies are given as $\omega _{m}=Nmv/r_{e}$,
$m=1,2,\ldots $, i.e., only the harmonics $k=Nm$ are radiated with the power
$N^{2}$ times larger than the corresponding radiation for a single particle.

First, we consider the case of a single particle with the angular coordinate
$\varphi =vt/r_{e}$. In accordance with the problem symmetry, for the
corresponding electric field we can write the Fourier expansion \cite%
{Grig06c,Arzu10} $\mathbf{E}(r,\theta ,\varphi ,t)=\sum_{s=-\infty
}^{+\infty }\mathbf{E}_{s}(r,\theta )e^{is(\varphi -vt/r_{e})}$, where $%
\mathbf{E}_{-s}(r,\theta )=\mathbf{E}_{s}^{\ast }(r,\theta )$. Ignoring the
absorption in the material of the ball, the total radiation intensity can be
evaluated in terms of the work done by the electromagnetic field on the
charge:%
\begin{equation}
I_{1}=-q\mathbf{v}\cdot \mathbf{E}(r_{e},\pi
/2,vt/r_{e})=-2qv\sum_{s=1}^{\infty }\mathrm{Re}\,[E_{s,\varphi }(r_{e},\pi
/2)],  \label{I1}
\end{equation}%
where $E_{s,\varphi }(r_{e},\theta )$ is the azimuthal component of the
Fourier coefficient. For a chain of point charges with the locations $%
\varphi =\varphi _{j}+vt/r_{e}$, the corresponding Fourier expansion reads%
\begin{equation}
\mathbf{E}^{(N)}(r,\theta ,\varphi ,t)=\sum_{j^{\prime
}=1}^{N}\sum_{s=-\infty }^{+\infty }\mathbf{E}_{s}(r,\theta )\exp \left[
is(\varphi -\varphi _{j^{\prime }}-vt/r_{e})\right] .  \label{EN}
\end{equation}%
The energy radiated per unit time by the $j$-th particle in the chain is
expressed as%
\begin{equation}
I_{j}^{(N)}=-2qv\sum_{j^{\prime }=1}^{N}\sum_{s=1}^{\infty }\mathrm{Re}%
\,[E_{s,\varphi }(r_{e},\pi /2)e^{is(\varphi _{j}-\varphi _{j^{\prime }})}],
\label{IjN}
\end{equation}%
and for the radiation intensity from the chain we get $I^{(N)}=-2qv%
\sum_{s=1}^{\infty }F_{N}(s)\mathrm{Re}\,[E_{s,\varphi }(r_{e},\pi /2)]$,
where the structure factor is given by (\ref{FN}). In the case of
equidistant distribution with $\varphi _{j}=2\pi (j-1)/N$, from (\ref{IjN})
one finds $I_{j}^{(N)}=-2Nqv\sum_{m=1}^{\infty }\mathrm{Re}\,[E_{Nm,\varphi
}(r_{e},\pi /2)]$ and this quantity is the same for all particles in the
chain. As a consequence, for the total radiation intensity from the
equidistant chain we get $I^{(N)}=-2N^{2}qv\sum_{m=1}^{\infty }\mathrm{Re}%
\,[E_{Nm,\varphi }(r_{e},\pi /2)]$. As seen from (\ref{EN}), for an
equidistant distribution and for harmonics $k=|s|=Nm$ the electric fields of
the charges with $j^{\prime }\neq j$ at the location of the $j$-th charge
are added coherently and the radiation from a single charge in the chain is
increased by $N$ times. Another factor in enhancement of the total radiation
comes from the number of particles in the chain.

Similar effect of coherent superposition of radiations from separate sources
will also be present if instead of separate charges we take bunches of
electrons. Note that in the problem with quasiequidistant distribution of
the bunches on a circle one can have two types of coherence. The first one
is the coherence between separate bunches and the second one corresponds to
the coherent effects in the radiation of a separate bunch. The coherent
synchrotron radiation (CSR) from a separate bunch has been widely
investigated in the literature both theoretically and experimentally (see,
for example, \cite{Naka89}-\cite{Evai19} and reviews \cite%
{Afan04,Hart02,Mull15}). For a single bunch, CSR is emitted in the spectral
range where the radiation wavelength is larger than the bunch length. It has
been observed in relatively wide range of wavelengths, from microwaves to
far infrared region. In many cases the theoretical results obtained within
the framework of the simplest model of a linear bunch with a vanishing
transverse size are in good agreement with observational data. More
complicated 2D and 3D models have been developed as well (see, for example,
\cite{Huan13,Stup21,Stup22} and references therein).

In the problem under consideration the radiation intensity for a single
linear bunch on a circular orbit can be obtained from the results given
above. The angular density of the number of quanta radiated by a single
bunch, $n_{N_{\mathrm{b}}}^{\mathrm{(b)}}(k,\theta )$, is given by the
formula $n_{N_{\mathrm{b}}}^{\mathrm{(b)}}(k,\theta )=F_{N_{\mathrm{b}}}^{%
\mathrm{(b)}}(k)n_{1}(k,\theta )$ (compare with (\ref{nN})), with $N_{%
\mathrm{b}}$ being the number of particles in the bunch. Similar to (\ref{FN}%
), for the bunch factor one has $F_{N_{\mathrm{b}}}^{\mathrm{(b)}%
}(k)=|\sum_{j=1}^{N_{\mathrm{b}}}\exp (-ik\varphi _{j}^{\mathrm{(b)}})|^{2}$%
, where $\varphi _{j}^{\mathrm{(b)}}$ determines the angular location of the
$j$-th particle in the bunch at $t=0$. Assuming that the coordinates of
separate particles are independent random variables we introduce the
distribution function $f(\varphi )$. The product $f(\varphi )d\varphi $ is
the probability to find an electron of the bunch in the angular interval $%
(\varphi ,\varphi +d\varphi )$. After averaging over the positions of a
charge in the bunch we find $F_{N_{\mathrm{b}}}^{\mathrm{(b)}}(k)=N_{\mathrm{%
b}}\left[ 1+(N_{\mathrm{b}}-1)g^{\mathrm{(b)}}(k)\right] $, where the bunch
form factor is given by $g^{\mathrm{(b)}}(k)=|\int_{0}^{2\pi }d\varphi
\,f(\varphi )e^{-ik\varphi }|^{2}$. Having this result, for the radiation of
a chain of bunches one gets $n_{N}(k,\theta )=F_{N}(k)F_{N_{\mathrm{b}}}^{%
\mathrm{(b)}}(k)n_{1}(k,\theta )$. Note that CSR in free space radiated by a
chain of electron bunches in submillimeter and millimeter wavelength range
has been observed in \cite{Ishi91,Shib91,Shib91b}.

\section{Conclusion}

\label{sec:Conc}

The coherence effects may essentially increase the radiation intensity
emitted by a system of charged particles. In the present paper we have
discussed those effects for the radiation from a chain of particles on a
circular orbit around a dielectric ball. In our previous research it has
already been demonstrated that in the problem with the same geometry and for
a single charge, under specific conditions on the parameters the influence
of the ball may increase the radiation intensity on a given harmonic by
orders of magnitude, compared with the radiation from the same charge
rotating in free space or in a transparent homogeneous medium. It has been
argued that the increase of the radiation intensity is related to the
constructive interference of the electromagnetic oscillations of the CR
generated inside the ball near the entire trajectory of the particle and
partially confined inside the ball by multiple reflections from its surface
\cite{Grig06}. The radiation is propagating in the angular range $\lesssim
\theta _{\mathrm{Ch}}$ near the rotation plane. Here we have provided an
analytical procedure for the determination of the location of the peaks in
the radiation intensity as a function of the ratio $r_{b}/r_{e}$.

For a chain of equidistant charges rotating around the ball, the
constructive interference of the waves generated by separate charges in the
chain gives rise to an additional increase in the radiation intensity by the
factor $N^{2}$ with $N$ being the number of particles in the chain. The
corresponding waves are radiated on the harmonics $k=mN$, $m=1,2,\ldots $.
We have shown that the relative shifts in the particles locations up to 10\%
do not destroy the coherence properties of the radiation. One can have
coherent radiation from a chain where the distribution of the particles is
not equidistant. An example is provided by the distribution (\ref{dist2})
where $r_{j}$ are rational numbers. In this case, radiation amplification by
the factor $N^{2}$ also takes place on certain harmonics $k=mk_{0}$.
However, unlike the case of equidistant distribution, in general, the
radiation on harmonics $k\neq mk_{0}$ is present as well. Similar coherence
effects will be present in the radiation from a chain of bunches circulating
around the ball. This phenomenon can be utilised in developing high-power
monochromatic sources of electromagnetic radiation in the GHz/THz frequency
ranges. We have demonstrated that considering the radiation from a chain
rotating around balls made of fused quartz and Teflon. The choice of those
materials is motivated by weak absorbtion of the radiation in the frequency
range under consideration. The latter is an essential point to have multiple
circulations of CR inside the ball. The radiation frequency can be
controlled by tuning the parameters $r_{e}$, $r_{b}$ or the energy of
particles in the chain. We have argued that similar coherent effects will
also appear in the radiation from a chain of bunches. In the special case of
an equidistant distribution of particles in the chain the results of the
present paper for the angle integrated frequency distribution of the
radiation are in agreement with those previously obtained in \cite{Arzu12}
for a dielectric ball made os fused quartz. Another special case with a
single charge circulating around a ball has been recently considered in \cite%
{Grig20}.

\section*{Acknowledgments}

The work was supported by the Science Committee of RA, in the frames of the
research project No. 21AG-1C069.

\end{document}